\documentstyle[pra,aps]{revtex}

\begin{document}
\title{Qubits as Parafermions}
\author{L.-A. Wu and D.A. Lidar}
\address{Chemical Physics Theory Group, University of Toronto, 80 St. George
Str., Toronto, Ontario M5S 3H6, Canada}
\maketitle

\begin{abstract}
Qubits are neither fermions nor bosons. A Fock space description of qubits
leads to a mapping from qubits to parafermions: particles with a hybrid
boson-fermion quantum statistics. We study this mapping in detail, and use
it to provide a classification of the algebras of operators acting on
qubits. These algebras in turn classify the universality of different
classes of physically relevant qubit-qubit interaction Hamiltonians. The
mapping is further used to elucidate the connections between qubits, bosons,
and fermions. These connections allow us to share universality results
between the different particle types. Finally, we use the mapping to study
the quantum computational power of certain anisotropic exchange
Hamiltonians. In particular, we prove that the XY model with
nearest-neighbor interactions only is {\em not} computationally universal.
We also generalize previous results about universal quantum computation with
encoded qubits to codes with higher rates.
\end{abstract}

\section{Introduction}

It is an experimental fact that there are only two types of {\em fundamental 
} particles in nature: bosons and fermions. Bosons are particles whose
wavefunction is unchanged under permutation of two identical particles. The
wavefunction of fermions is multiplied by $-1$ under the same operation. An
equivalent statement is that bosons transform according to the $1$-dimensional, symmetric, irreducible representation (irrep) of the
permutation group, while fermions belong to the $1$-dimensional
antisymmetric irrep. The permutation group has only these two $1$-dimensional irreps. What about particles transforming according to
higher-dimensional irreps of the symmetric group? Much research went into
studying this possibility, in the early days of the quark model, before the
concept of ``colored'' quarks gained widespread acceptance \cite
{Green:53,Greenberg:64}. However, there are now good reasons to believe that
particles obeying such ``parastatistics'' do not exist
(Ref.~\onlinecite[p.137]{Peres:book}). Nevertheless, as we will show
below, the traditional definition of 
a Hilbert space of qubits is inconsistent with the properties of either
bosons or fermions.

The description of bosons and fermions in terms of their properties under
particle permutations uses the language of first-quantization. A useful
alternative description is the second-quantized formalism of Fock space \cite
{Peres:book,March:book}. A basis state in the boson or fermion Hilbert-Fock
space can be written as $|n_{1}^{\alpha },n_{2}^{\alpha },...\rangle $,
where $ n_{i}^{\alpha }$ counts how many bosons ($\alpha =b$) or fermions ($
\alpha =f $) occupy a given mode, or site $i$. Note that the total number of
modes does not need to be specified in the Fock-basis. Ignoring
normalization, raising, $\alpha _{i}^{\dagger }$ (lowering, $\alpha _{i}$)
operators increase (decrease) $n_{i}^{\alpha }$ by $1$. A consequence of the
permutation properties of bosons and fermions is that their corresponding
raising and lowering operators satisfy commutation and anti-commutation
relations: 
\begin{eqnarray*}
\lbrack b_{i}^{\dagger },b_{j}^{\dagger }] &=&0,\quad \lbrack
b_{i},b_{j}^{\dagger }]=\delta _{ij}\qquad {\rm bosons} \\
\{f_{i}^{\dagger },f_{j}^{\dagger }\} &=&0,\quad \{f_{i},f_{j}^{\dagger
}\}=\delta _{ij}\qquad {\rm fermions.}
\end{eqnarray*}
From this follow a number of well-known facts \cite{Peres:book,March:book}.
Let $\hat{n}_{i}^{\alpha }=\alpha _{i}^{\dagger }\alpha _{i}$; this is the
number operator, which is diagonal in the Fock-basis $|n_{1}^{\alpha
},n_{2}^{\alpha },...\rangle $, and has eigenvalues $n_{i}^{\alpha }$. Then:

\begin{itemize}
\item $[b_{i}^{\dagger },b_{j}^{\dagger }]=0$ $\Longrightarrow $ an
arbitrary number of bosons $n_{i}^{\alpha }$ can occupy a given mode $i$. On
the other hand, $\{f_{i}^{\dagger },f_{j}^{\dagger }\}=0$ $\Longrightarrow $
only $n_{i}^{f}=0,1$ is possible for fermions.

\item $[b_{i},b_{j}^{\dagger }]=\delta _{ij}\Longrightarrow $ the Hilbert
space of bosons has a natural tensor product structure, i.e., $
|n_{1}^{b},n_{2}^{b},...\rangle =|n_{1}^{b}\rangle \otimes |n_{2}^{b}\rangle
\otimes \cdots $. More specifically, it is possible to {\em independently}
operate on each factor of the Hilbert space. However, 
\begin{eqnarray*}
\{f_{i},f_{j}^{ \dagger }\} = \delta _{ij}\Longrightarrow
f_{j}|n_{1}^{f},...,n_{j-1}^{f},1,n_{j+1}^{f},...\rangle = (-1)^{
\sum_{k=1}^{j-1}n_{k}^{f}}|n_{1}^{f},...,n_{j-1}^{f},0,n_{j+1}^{f},...
\rangle,
\end{eqnarray*}
which means that the outcome of operating on a mode of a multi-fermion state
depends on all previous modes (the order of modes is actually arbitrary).
This non-local property means that the fermionic Fock space does not have a
natural tensor product structure, although it can be mapped onto one that
does using the Jordan-Wigner transformation \cite{Jordan:28} (see \cite
{Bravyi:00} for a more detailed discussion).
\end{itemize}

What about qubits? The standard notion of what a qubit is, is the following 
\cite{Nielsen:book}:

\noindent {\bf Qubit}:

\begin{itemize}
\item A qubit is a vector in a two-dimensional Hilbert space ${\cal H}_{i}= 
{\rm span}\{|0\rangle _{i},|1\rangle _{i}\}$ (like a fermion), and

\item An $N$-qubit Hilbert space has a tensor product structure: ${\cal H}
=\otimes _{i=1}^{N}{\cal H}_{i}$ (like bosons).
\end{itemize}

It appears that a qubit is a hybrid fermion-boson particle! We conclude that 
{\em qubits do not exist as fundamental particles}. This motivates us to
consider an intermediate statistics of ``parafermions'' in order to have a
Fock space description of a qubit. We define the parafermionic commutation
relations by \cite{comment1}: 
\begin{eqnarray}
\{a_{i},a_{i}^{\dagger }\} &=&1,  \nonumber \\
\lbrack a_{i},a_{j}^{\dagger }] &=&0\quad {\rm if\quad }i\neq j.
\label{eq:pfcomm}
\end{eqnarray}
Here $i,j$ are different modes, or different qubits. The relation $
[a_{i},a_{j}^{\dagger }]=0$ for $i\neq j$ immediately implies a tensor
product structure, while $\{a_{i},a_{i}^{\dagger }\}=1$, which together with 
$a_{i}|{\tt 0}\rangle =0$ ($\left| {\tt 0}\right\rangle $ is the vacuum
state) implies 
\begin{equation}
a_{i}a_{i}=a_{i}^{\dagger }a_{i}^{\dagger }=0  \label{eq:aa=0}
\end{equation}
in the standard (irreducible) two-dimensional representation. Therefore a
double-occupation state cannot be realized, i.e., the single-particle
Hilbert space is two-dimensional. These are exactly the requirements for a
qubit.

In fact, the notion of particles with ``intermediate'' statistics such as
parafermions is well known and established in condensed matter physics,
e.g., hard-core bosons, excitons, or the Cooper pairs of superconductivity 
\cite{Mahan:book} (see also Section \ref{making}). Such particles are always 
{\em composite}, i.e., they are not fundamental. Another way of obtaining a
particle that is neither a boson nor a fermion is to simply ignore one or
more degrees of freedom. This is by and large the approach taken in current
proposals for the physical implementation of quantum computers. For example,
a single spin-$\frac{1}{2}$, without the orbital component of its
wavefunction, behaves exactly like a qubit. This is the case of the
electron-spin qubit in quantum dots \cite{Loss:98}. Related to this, a
truncated multi-level atom can also approximate a qubit, as in the ion-trap
proposal \cite{Cirac:95}. What are the implications of this for quantum
computing (QC)? In a nutshell, ``ideal'' qubits are hard to come by. If a
qubit is to exist as an approximate two-level system, or as a composite
particle, or as a partial description of an object with additional degrees
of freedom, this means that some robustness is lost and the door is opened
to decoherence. E.g., the additional levels in a multi-level Hilbert space
can cause ``leakage'', the orbital degrees of freedom act as a bath coupled
to the spin-qubit, and a composite particle may decay (e.g., the
exciton-qubit \cite{Chen:01}).

The advantages of the parafermionic formalism for qubits, however, are not
necessarily in understanding these sources of decoherence, because this
formalism ``accepts'' qubits as particles. Instead, the parafermionic
formalism allows us to naturally establish mappings between qubits,
fermions, and bosons. This mapping serves to transport well-known results
about one type of particle to another, which, as we show below, clarifies
questions regarding the ability of sets of one type of particle to act as
universal simulators \cite{Lloyd:96} of sets of another type of particle. It
also helps in connecting the Hamiltonians of condensed matter physics to
standard tools of quantum computation.

The structure of the paper is as follows. In the next section we formally
introduce the second quantization of qubits. We then classify the algebraic
structure of parafemionic operators in Section \ref{subalgebras}. This
classification, into subalgebras with different conservation properties, is
very useful for establishing which subsets of qubit operators are universal,
either on the full Hilbert space, or only on a subspace. This is taken up in
the next two sections, where we establish the connection between
parafermions and fermions (Section \ref{fermpara}) and bosons (Section \ref{bosepara}). The connection to fermions and bosons also works in the
opposite direction: we are able to classify which fermionic and bosonic
operator sets are universal. This has implications, e.g., for the linear
optics quantum computing proposals \cite{Knill:00,Ralph:01}. Section \ref{making} shows how to construct parafermions out of paired fermions and
bosons, emphasizing the compound-particle aspect of qubits. With the
connections between fermions, bosons, and parafermions clarified, we explain
in Section \ref{bilinuniv} a remarkable difference between parafermions and
the other particle types: bilinear parafermionic Hamiltonians are sufficient
for universal quantum computation, whereas fermionic and bosonic
Hamiltonians are not. In Section \ref{fluctuations} we briefly use the
mapping to fermions to derive the thermal fluctuations of non-interacting
parafermions at finite temperature. In Section \ref{universality} we apply
the classification of the various parafermionic operator subalgebras to the
problem of establishing universality of typical Hamiltonians encountered in
solid state physics. We generalize a number of our previous results \cite{WuLidar:01,LidarWu:01}. In particular, we establish that the XY model is
not universal with nearest-neighbor interactions only; and, we prove
universality of the XXZ model for codes with arbitrarily high rates. We
conclude in Section \ref{conclusions}.

\section{Second Quantization of Qubits}

\label{secquan}

As in the cases of bosons and fermions, a parafermion number operator in
mode $i$ can be defined as

\[
\hat{n}_{i}=a_{i}^{\dagger }a_{i}, 
\]
with eigenvalues $n_{i}=0,1$. The total number operator is $\hat{n}=\sum_{i} 
\hat{n}_{i}$. A normalized basis state in the parafermionic Fock space is 
\[
|\cdots n_{i}\cdots \rangle =\prod_{i}(a_{i}^{\dagger })^{n_{i}}\left| {\tt 
0 }\right\rangle , 
\]
which we think of as representing a state with the $i^{{\rm th}}$ qubit in
the ``up'' (``down'') state if the $i^{{\rm th}}$ parafermion is present
(absent), i.e., $n_{i}=1$ ($0$). {\em Qubit} {\em computational basis states
are thus mapped to parafermionic Fock states}.{\em \ }Equivalently, consider
the following mapping from qubits to parafermions: 
\begin{eqnarray*}
\left| 0_{1}\cdots 0_{i-1}0_{i}0_{i+1}\cdots \right\rangle &\rightarrow
&\left| {\tt 0}\right\rangle \\
\left| 0_{1}\cdots 0_{i-1}1_{i}0_{i+1}\cdots \right\rangle &\rightarrow
&a_{i}^{\dagger }\left| {\tt 0}\right\rangle ,
\end{eqnarray*}
where on the left $0$ and $1$ represent the standard (first-quantized)
logical states of a qubit. {\em Qubits can thus be identified with
parafermionic operators}.

The mapping of qubits to parafermions is completed by mapping the Pauli
matrices $\sigma _{i}^{\alpha }$ to parafermionic operators:

\begin{equation}
\sigma _{i}^{+}\rightarrow a_{i}^{\dagger }\qquad \sigma _{i}^{-}\rightarrow
a_{i}\qquad \sigma _{i}^{z}\rightarrow 2n_{i}-1.  \label{eq:sl2}
\end{equation}
It is then straightforward to check that the standard $sl(2)$ commutation
relations of the Pauli matrices, 
\begin{eqnarray*}
\lbrack \sigma _{i}^{+},\sigma _{j}^{-}] &=&2\delta _{ij}\sigma _{i}^{z} \\
\lbrack \sigma _{i}^{z},\sigma _{j}^{\pm }] &=&\pm \delta _{ij}\sigma
_{i}^{\pm },
\end{eqnarray*}
are preserved, so that we have a faithful second quantized representation of
the qubit system Hilbert space and algebra
(Of course we could also have mapped $su(2)=\{\sigma ^{x},\sigma ^{y},\sigma
^{z}\}$ to the parafermionic operators, by appropriate linear combinations.)
To illustrate the multi-qubit Hilbert-Fock space representation, consider
the case of two modes, i.e., $i,j=1,2$. The space splits into a vacuum state 
$|00\rangle =\left| {\tt 0}\right\rangle $, single-particle states $
|01\rangle =a_{1}^{\dagger }\left| {\tt 0}\right\rangle $ and $|10\rangle
=a_{2}^{\dagger }\left| {\tt 0}\right\rangle $, and a two-particle state $
|11\rangle =a_{1}^{\dagger }a_{2}^{\dagger }\left| {\tt 0}\right\rangle $.
It is important to emphasize that the parafermionic formalism is
mathematically equivalent to the standard Pauli matrix formalism. We will be
using both in the sections below, starting with the parafermionic, as it
makes particularly transparent the translation of known results about
fermions to qubits.

\section{General Properties of Parafermionic Operators}

\label{subalgebras}

$N$-qubit operators in QC are elements of the group $U(2^{N})$. We will
begin our discussion by identifying a set of infinitesimal parafermionic
generators for $U(2^{N})$. Recall that with any $r$-parameter Lie group
there are associated $r$ infinitesimal generators \cite{Wybourne:book}.
E.g., in the case of $su(2)$ these are, in the two-dimensional irreducible
representation, the Pauli matrices $\{\sigma _{x},\sigma _{y},\sigma _{z}\}$. Now, let $\alpha =\{\alpha _{i}\},\beta =\{\beta _{j}\}$, where $\alpha
_{i}$,$\beta _{j}$ can be $0$ or $1$. In terms of parafermionic operations,
any element of $U(2^{N})$ can be written as $U(b)=\exp (-i\sum_{\alpha,\beta
}b^{\alpha \beta }Q_{\alpha ,\beta }(N))$, where $b^{\alpha \beta }$ are
continuous parameters (generalized Euler angles) and the \ $2^{N}\times 2^{N}
$ {\em infinitesimal group generators} $Q_{\alpha ,\beta }(N)$ are defined
as follows:\ let $N_{\alpha }=\sum_{i=1}^{N}\alpha _{i}$, and 
\begin{equation}
q_{\alpha }^{\dagger }(N_{\alpha })=(a_{N}^{\dagger })^{\alpha _{N}}\cdots
(a_{1}^{\dagger })^{\alpha _{1}},\quad q_{\beta }(N-N_{\alpha
})=a_{N}^{\beta _{N}}\cdots a_{1}^{\beta _{1}}.
\end{equation}
Then:

\begin{equation}
Q_{\alpha ,\beta }(N)=q_{\alpha }^{\dagger }(N_{\alpha })q_{\beta
}(N-N_{\alpha }).
\end{equation}
The $Q_{\alpha ,\beta }(N)$ will be recognized as all possible
transformations between $N$-qubit computational basis states,
e.g., for $N=2$ the set of $16$ operators is:\newline
\noindent $ \{I,a_{1}^{\dagger }, a_{2}^{\dagger }, a_{1}, a_{2},
a_{2}^{\dagger }a_{1}^{\dagger }, a_{1}a_{2}, a_{1}^{\dagger
}a_{1},a_{1}^{\dagger }a_{2}, a_{2}^{\dagger }a_{1},a_{2}^{\dagger }a_{2},
a_{2}^{\dagger }a_{1}^{\dagger }a_{1}, a_{2}^{\dagger }a_{1}^{\dagger
}a_{2}, a_{1}^{\dagger }a_{1}a_{2}, a_{2}^{\dagger }a_{2}a_{1},
a_{2}^{\dagger }a_{1}^{\dagger }a_{2}a_{1}\}$, where $I$ is the identity
operator. The set $ Q_{\alpha ,0}(N)$ generates all possible basis
states from the vacuum state. Hermitian forms are $Q+Q^{\dagger }$ and $
i(Q-Q^{\dagger })$. We will turn to the hermitian set of generators in the
discussion of applications, in Section \ref{universality}.

Note that infinitesimal generators are not the generators one usually
considers in QC. Rather, in QC, a gate operation is obtained by the unitary
evolution generated through the turning on/off of a set of physically
available {\em Hamiltonians} $\{H_{\mu }\}$, that are generally a small
subset of the $2^{N}\times 2^{N}$ infinitesimal generators $Q_{\alpha ,\beta
}(N)$. ``Generated'' here has the usual meaning of allowing linear
combinations and commutation of Hamiltonians. We will say that {\em a set of
Hamiltonians }$\{H_{\mu }\}${\em is universal with respect to a Lie group} $
{\cal G}$ if it generates the Lie algebra of that group. The question of the
dimension of the universal set of Hamiltonians with respect to $U(2^{N})$ is
somewhat subtle, since it is context dependent. Lloyd showed that given two
non-commuting operators $A$,$B$, represented by $n\times n$ matrices, one
can almost always generate $U(n)$ \cite{Lloyd:95}. However, it is not
necessarily clear how this result is related to {\em physically available}
Hamiltonians, since in practice one may have only limited control over terms
in a Hamiltonian. E.g., the standard Hamiltonian generators for $SU(4)$ (two
qubits) is the $5$-element set $\{\sigma _{1}^{z},\sigma _{2}^{z},\sigma
_{1}^{x},\sigma _{2}^{x},\sigma _{1}^{z}\sigma _{2}^{z}\}$. However, the $4$
-element set $\{\sigma _{1}^{z},\sigma _{2}^{z},\sigma _{1}^{z}\sigma
_{2}^{x}-\sigma _{1}^{x}\sigma _{2}^{z},{\vec{\sigma}}_{1}\cdot {\vec{\sigma}
}_{2}\}$ also generates $SU(4)$, and may be physically available \cite{WuLidar:01}. Another example are the following sets of, respectively, five,
four, and three generators: $\{\sigma _{1}^{x},\sigma _{2}^{x},\sigma
_{1}^{z},\sigma _{2}^{z},\sigma _{1}^{z}\sigma _{2}^{z}\},\{\sigma
_{1}^{x},\sigma _{2}^{x},c_{1}\sigma _{1}^{z}+c_{2}\sigma _{2}^{z},\sigma
_{1}^{z}\sigma _{2}^{z}\},\{\sigma _{1}^{x},\sigma _{2}^{x},c_{1}\sigma
_{1}^{z}+c_{2}\sigma _{2}^{z}+c_{3}\sigma _{1}^{z}\sigma _{2}^{z}\}$ (where $
c_{i}$ are constants). Which set of generators is physically available
(i.e., directly controllable) depends on the specific system used to
implement the quantum computer. As we will show below, it is sometimes the
case that a given, physically available, set of Hamiltonians is universal
with respect to a {\em subgroup} of $U(2^{N})$, which may be quite useful,
provided the subgroup is sufficiently large (typically, still exponential in $N$). This notion of universality with respect to a subgroup is what gives
rise to the idea of {\em encoded universality} \cite{Bacon:99a,DiVincenzo:00a,Bacon:Sydney,Lidar:00b}: one encodes a logical
qubit into two or more physical qubits, and studies the universality of the
subgroup-generating Hamiltonians with respect to these encoded/logical
qubits.

The infinitesimal parafermionic generator $Q_{\alpha ,\beta }(N)$ can be
rearranged into certain subsets of operators with clear physical meaning,
which we now detail.

\begin{enumerate}
\item Local subalgebras: The tensor product structure of qubits is naturally
enforced by $[a_{i},a_{j}^{\dagger }]=0$ for $i\neq j$. This induces a
tensor product structure $\bigotimes_{i=1}^{N}sl_{i}(2)$ on the subalgebras
formed by the grouping $sl_{i}(2)=\{a_{i},a_{i}^{\dagger },1-2n_{i}\}$. Each 
$sl_{i}(2)$ can only change states within the same mode.

\item SA$p$ -- Subalgebra with {\em conserved parity}: Define a {\em parity }
operator as 
\[
\hat{p}=(-1)^{\hat{n}}. 
\]
It has eigenvalues $1$ ($-1$) for even (odd) total particle number. The
operators that commute with the parity operator form a subalgebra, which we
denote by SA$p$. Let $k$ ($l$) be the number of $a_{i}^{\dagger }$ ($a_{i}$)
factors in $Q_{\alpha ,\beta }(N)$, i.e., 
\[
k=\sum \alpha _{i},\quad l=\sum \beta _{i}. 
\]
SA$p$ consists of those operators having $k-l$ even, so its dimension (i.e.,
number of generators) is $2^{2N}/2$. To see this, let $Q_{I}$ be in SA$p$,
and consider its action on a state with an even number of particles $
|n\rangle $. Since $k-l$ is even, $Q_{I}|n\rangle =|n^{\prime }\rangle $
where $n^{\prime }$ is also even. Now, $\hat{p}Q_{I}|n\rangle =\hat{p}
|n^{\prime }\rangle =+|n^{\prime }\rangle $, but also $Q_{I}\hat{p}|n\rangle
=Q_{I}(+|n\rangle )=|n^{\prime }\rangle $ so $[\hat{p},Q]=0$,
e.g., for $N=2$ SA$p$ consists of: $ \{I, a_{2}^{\dagger }a_{1}^{\dagger },
a_{1}a_{2}, a_{1}^{\dagger }a_{1}, a_{1}^{\dagger }a_{2}, a_{2}^{\dagger
}a_{1}, a_{2}^{\dagger }a_{2}, a_{2}^{\dagger }a_{1}^{\dagger }a_{2}a_{1}\}$.

\item SA$n$ -- Subalgebra with {\em conserved particle number}. This
subalgebra, which we denote SA$n$, is formed by all operators commuting with
the number operator $\hat{n}$. These are the operators for which $k=l$, so
its dimension is $\sum_{k=0}^{N}{\ {
{{N}  \choose {k}}
} }^{2}=\frac{(2N)!}{N!N!}$. To see this, let $Q_{II}$ be in SA$n$, and
consider its action on a state $|n\rangle $ with $n$ particles. $Q_{II}$
cannot change this number since $k=l$, but it can transform $|n\rangle $: $ 
\hat{n}Q_{II}|n\rangle =\hat{n}|n\rangle ^{\prime }=n|n\rangle ^{\prime }$.
However, $Q_{II}\hat{n}|n\rangle =nQ_{II}|n\rangle =n|n\rangle ^{\prime }$,
so $[Q_{II},\hat{n}]=0$,
e.g., for $N=2$ SA$n$ consists of: $ \{I, a_{1}^{\dagger }a_{1},
a_{1}^{\dagger }a_{2}, a_{2}^{\dagger }a_{1}, a_{2}^{\dagger }a_{2},
a_{2}^{\dagger }a_{1}^{\dagger }a_{2}a_{1}\}$. Clearly, SA$n\subset$SA$p$.

\item Subsets of bilinear operators: There are two types of bilinear
operators for $i\neq j$: $a_{i}^{\dagger }a_{j}$ (which conserve the
particle number), and $a_{i}a_{j},a_{i}^{\dagger }a_{j}^{\dagger }$ (which
conserve parity). Let $\mu =(ij)$, then first: 
\begin{eqnarray}
T_{\mu }^{x} &=&a_{j}^{\dagger }a_{i}+a_{i}^{\dagger }a_{j}  \nonumber \\
T_{\mu }^{z} &=&n_{i}-n_{j}  \label{eq:slt}
\end{eqnarray}
and $T_{\mu }^{y}=i[T_{\mu }^{x},T_{\mu }^{z}]$ form an $su(2)$ subalgebra,
denoted $su_{\mu }^{t}(2)$. Clearly, $su_{\mu }^{t}(2)\in $SA$n$. Second: 
\begin{eqnarray}
R_{\mu }^{x} &=&a_{i}a_{j}+a_{i}^{\dagger }a_{j}^{\dagger }  \nonumber \\
R_{\mu }^{z} &=&n_{i}+n_{j}-1  \label{eq:slr}
\end{eqnarray}
and $R_{\mu }^{y}$\ form another $su(2)$ subalgebra, denoted $su_{\mu
}^{r}(2)\in $SA$p$. Note that $[su_{\mu }^{t}(2),su_{\mu }^{r}(2)]=0$ since
any product of raising/lowering operators from these algebras contains a
factor of $a_{i}a_{i}$ or $a_{i}^{\dagger }a_{i}^{\dagger }$. Consider as an
example the case of $N=2$ modes. Whereas the direct product group $
SU_{1}(2)\otimes SU_{2}(2)$ yields all product states, the group $
SU^{t}(2)\oplus SU^{r}(2)$ can transform between states with equal particle
number and states differing by two particle numbers.

\item Generators of SA$n$($N)$: The set of Hamiltonians $\{a_{i}^{\dagger
}a_{j}\}_{i,j=1}^{N+1}$ generates SA$n$($N$), i.e., the subalgebra of
conserved particle number on $N$ modes (qubits). Proof: this set maps to the 
XY model (see Section \ref{XY}). The rest follows using the method of \cite
{LidarWu:01}. Note that $ \{a_{i}^{\dagger }a_{j}\}_{i,j=1}^{N}$ does {\em 
not} generate SA$n$($N+1$), since this set cannot generate $\hat{n}_{1}\hat{n
}_{2}\cdots \hat{n}_{N}$.

\item Generators of SA$p$($N)$: The set of Hamiltonians $\{a_{i}^{\dagger
}a_{j},a_{i}a_{j}+a_{i}^{\dagger }a_{j}^{\dagger
},i(a_{i}a_{j}-a_{i}^{\dagger }a_{j}^{\dagger })\}_{i,j=1}^{N}$ yield all
states with even particle number on $N$ modes from the vacuum state. (Proof
is trivial.)

\item Generators of $SU(2^{N})$: In order to transform between states
differing by an odd number of particles it is necessary to include the
operators $\{a_{i},a_{i}^{\dagger }\}$ as well. The corresponding set $
\{a_{i}^{\dagger }a_{j},a_{i}a_{j},a_{i}^{\dagger }a_{j}^{\dagger
},a_{i},a_{i}^{\dagger }\}_{i,j=1}^{N}$ generates a set of universal gates
(proof is trivial), and then by standard universality results \cite
{Barenco:95a,DiVincenzo:95} the entire $SU(2^{N})$.
\end{enumerate}

Additional structure emerges from a mapping between fermions and
parafermions. This structure can help us both in simulating fermionic system
using qubits, and the understanding of universality of qubit systems.

\section{Fermions and Parafermions}

\label{fermpara}

A general fermionic Fock state is

\begin{equation}
\left| n_{1},n_{2},\cdots \right\rangle _{F},  \label{eq:fermion}
\end{equation}
where $n_{i}=0,1$ is the occupation number of mode $i$. As is well known 
\cite{Judd:book}, the fermionic (``supergroup'' \cite{Wybourne:book}) $
U(2^{N})$ has infinitesimal generators 
\[
\tilde{Q}_{\alpha ,\beta }^{f}(N)=(f_{N}^{\dagger })^{\alpha _{N}}\cdots
(f_{1}^{\dagger })^{\alpha _{1}}Af_{N}^{\beta _{N}}\cdots f_{1}^{\beta _{1}} 
\]
where 
\[
A=\bigotimes_{i=1}^{N}(1-n_{i}). 
\]
This basis is equivalent by a linear transformation to the more familiar set 
\[
Q_{\alpha ,\beta }^{f}(N)=(f_{N}^{\dagger })^{\alpha _{N}}\cdots
(f_{1}^{\dagger })^{\alpha _{1}}f_{N}^{\beta _{N}}\cdots f_{1}^{\beta _{1}} 
\]
which transforms between all possible fermionic Fock states (``fermionic
computational basis state''). There is a group chain of this group, 
\begin{equation}
U(2^{N})\supset SO(2N+1)\supset SO(2N)\supset U(N)  \label{eq:subchain}
\end{equation}
and the generators of the subgroups are known \cite{Wybourne:book}.

The Jordan-Wigner (JW) transformation \cite{Jordan:28}, recently generalized
in Ref.~\onlinecite{Batista:01}, allows one to establish an isomorphism
between fermions and parafermions. Defining 
\begin{equation}
S_{i}^{f}\equiv \bigotimes_{k=1}^{i-1}(1-2n_{k}^{f}),\quad S_{i}\equiv
\bigotimes_{k=1}^{i-1}(1-2n_{k}),
\end{equation}
the mapping is:

\begin{eqnarray}
n_{i}^{f} &\rightarrow &n_{i},  \nonumber \\
f_{i} &\rightarrow &a_{i}S_{i},  \nonumber \\
f_{i}^{\dagger } &\rightarrow &a_{i}^{\dagger }S_{i}.  \label{eq:JW}
\end{eqnarray}
The action of the fermionic operators on the state (\ref{eq:fermion}) is
equivalent to that of the corresponding parafermionic operators on the state 
$\left| n_{1},n_{2},\cdots \right\rangle $. To see this, note that $
[a_{i},S_{i}]=0$. Therefore the effect of the JW transformation is quite
simple: by commuting all $S_{i}$ to the left when when mapping a fermionic
infinitesimal generator to a parafermionic one, we see that (i) the
parafermionic $a_{i},a_{i}^{\dagger }$ operators will yield a state with the
same parafermionic occupation numbers as the corresponding fermionic state,
(ii) the action of the product of $S_{i}$'s is to produce a phase $\pm 1$.
(This may become a relative phase when acting on a state that is a
superposition of computational basis states.) This allows us to study
algebraic properties of one set of particles in terms of the other.

Using the JW transformation we find that the same subgroup chain (\ref
{eq:subchain}) holds for parafermions, and we can immediately write down
also the infinitesimal generators for the corresponding parafermionic
subgroups. The result is given in Table~\ref{tab1}.

The significance of these subgroups for QC is in the classification of the
universality properties of fermionic and parafermionic Hamiltonians. E.g., a
Hamiltonian of non-interacting fermions, i.e., one including only bilinear
terms $\{f_{i}^{\dagger }f_{j},f_{i}f_{j},f_{j}^{\dagger }f_{i}^{\dagger }\}$
is not by itself universal since it merely generates $SO(2N)$. Recent work
has clarified what needs to be added to such a Hamiltonian in order to
establish universality \cite{Bravyi:00,Terhal:01,Knill:01}. Regarding $
SO(2N+1)$, note that one must carefully discuss the hermitian terms $
f_{i}+f_{i}^{\dagger }$ and $i(f_{i}-f_{i}^{\dagger })$ if one wants to
consider them as Hamiltonians, since it is unclear which physical process
can be described by such Hamiltonians (a single fermion
creation/annihilation operator can turn an isolated fermion into a boson, a
process that does not seem to occur in nature).

A more powerful classification, from the QC viewpoint, is in terms of
physically available Hamiltonian generators of the subgroups. An interesting
restriction of the set of infinitesimal generators to a physically
reasonable set of Hamiltonians is to consider only nearest-neighbor
interactions, where possible. The results known to us in this case are
presented in Table~\ref{tab2}.

A couple of comments are in order regarding Table~\ref{tab2}: First, note
the group $ SO(2N+1)$ may be unphysical not just for fermions since its
generators must contain terms like $f_{i}+f_{i}^{\dagger }$ in its
Hamiltonian, but also for parafermions: it requires a non-local Hamiltonian
due to the $S_{i}$ term. Second, the corresponding fermionic generators for $
U(2^{N})$ given here is unphysical because it includes terms that are linear
in $f_{i}$ and furthermore non-local. A physically acceptable set is $
\{f_{i}^{\dagger }f_{i+1},f_{i}f_{i+1},f_{i}^{\dagger }f_{i+1}^{\dagger
}f_{i}f_{i+1},{\rm \ h.c.}\}$, but this set is not universal over the full $
2^{N}$-dimensional Hilbert space (since it conserves parity). This means
that a qubit needs to be encoded into two fermions in this case, a situation
we explore further in Section \ref{making}. Now let us verify the claims of
Table~\ref{tab2}. Our strategy is to show that in each case, we can use the
Hamiltonians for generating all infinitesimal generators of the
corresponding subgroup in Table~\ref{tab1}.

Consider first the subgroup $SU(N)$: In the fermionic case, we claim that
this subgroup has nearest neighbor Hamiltonian generators $f_{i}^{\dagger
}f_{i+1}$ and their hermitian conjugates. E.g., for $N=3$, if we have the
four operators $f_{1}^{\dagger }f_{2}$, $f_{2}^{\dagger }f_{3}$ and h.c.,
then we can generate $f_{1}^{\dagger }f_{3}=[f_{1}^{\dagger
}f_{2},f_{2}^{\dagger }f_{3}]$ and h.c., as well as $\hat{n}_{i}^{f}-\hat{n}
_{j}^{f}=[f_{i}^{\dagger }f_{j},f_{j}^{\dagger }f_{i}]$. This yields a total
of nine operators, eight of which are linearly independent, that generate $
SU(3)$. As for parafermions, we can use the JW transformation to get $
f_{i+1}^{\dagger }f_{i}\rightarrow a_{i+1}^{\dagger
}S_{i+1}a_{i}S_{i}=a_{i+1}^{\dagger }(1-2\hat{n}_{i})a_{i}=a_{i+1}^{\dagger
}a_{i}$ (where we have used $[a_{i},S_{i}]=0$ and $\hat{n}
_{i}a_{i}=a_{i}^{\dagger }a_{i}a_{i}=0$). This establishes an isomorphism
between the fermionic and parafermionic generators for $SU(N)$. Hence the
parafermionic subgroup $SU(N)$ is generated by $a_{i}^{\dagger }a_{i+1}$ and
h.c.

Now consider $SO(2N)$: In the fermionic case we have $f_{1}^{\dagger
}f_{2}^{\dagger }$, and using the result for $U(N)$ we also have $
f_{4}^{\dagger }f_{1}$; therefore we have $[f_{4}^{\dagger
}f_{1},f_{1}^{\dagger }f_{2}^{\dagger }]=$ $f_{4}^{\dagger }f_{2}^{\dagger }$.Clearly, the interaction range can be extended to cover all generators.
For the parafermionic case, using the JW transformation we find $
f_{i+1}^{\dagger }f_{i}^{\dagger }\rightarrow a_{i+1}^{\dagger
}S_{i+1}a_{i}^{\dagger }S_{i}=a_{i+1}^{\dagger }(1-2\hat{n}
_{i})a_{i}^{\dagger }=a_{i+1}^{\dagger }a_{i}^{\dagger }$, so that we again
have an isomorphism with the fermionic case.

Next consider the (unphysical)\ subgroup $SO(2N+1)$:\ In the fermionic case
it suffices to note that $\frac{1}{2}[f_{i},f_{j}]=f_{i}f_{j}$ and $\frac{1}{
2}[f_{i}^{\dagger },f_{j}]=f_{i}^{\dagger }f_{j}$, so that we can generate
all infinitesimal generators by the linear terms $f_{i}$ and $f_{i}^{\dagger}$. The parafermionic case follows by the JW-transformation.

Finally, in the $U(2^{N})$ case the universality of the parafermionic set $
\{a_{i},a_{i}^{\dagger }a_{i+1},{\rm h.c.}\}$ follows from that of the set
of all single qubit operations together with the Hamiltonian of the
nearest-neighbor XY model [Eq.~(\ref{eq:xy}) below], proved in Ref.~\onlinecite{Imamoglu:99}. The fermionic case follows by the
JW-transformation.

Let us recapitulate the meaning of the results presented in this section: we
have shown how to classify subalgebras of fermionic/parafermionic operators
in terms of the groups they generate. This therefore classifies their
universality properties with respect to these groups. This is particularly
important in the context of a given set of physically available
Hamiltonians. Our method employed a mapping between fermions and
parafermions, which allowed us to easily transport known results about one
type of particle to the other.

\section{Bosons from Parafermions}

\label{bosepara}

A linear combination of different-mode parafermions can approximately form a
boson. Define 
\[
B=\frac{1}{\sqrt{N}}\sum_{i=1}^{N}a_{i}. 
\]
Then using Eq.~(\ref{eq:pfcomm}) we have 
\[
\lbrack B,B^{\dagger }]=\frac{1}{N}\sum_{i=1}^{N}1-2\hat{n}_{i}=1-\frac{ 2 
\hat{n}}{N}. 
\]
If the parafermion number is much smaller than the available number of
sites/modes, i.e., when $n\ll N$, then $[B,B^{\dagger }]\approx 1$, which is
an approximate single-mode boson commutation relation.

To get $K$ boson modes, we can divide $N$ into $K$ approximately equal
parts. Each part has $N_{\alpha }=N/K$ qubits and approximately represents a
boson. The $k^{{\rm th}}$ boson is $B_{\alpha }=\frac{1}{\sqrt{N_{\alpha }}}
\sum_{i=1}^{N_{\alpha }}a_{i}$. Then 
\[
\lbrack B_{\alpha },B_{\beta }^{\dagger }]=\delta _{\alpha \beta }(1-\frac{2 
\hat{n}_{\alpha }}{N_{\alpha }})\stackrel{{n}_{\alpha }\ll N_{\alpha }}{
\approx }\delta _{\alpha \beta }. 
\]

Physically, this means that a low-energy qubit system (with most qubits in
their ground state) can macroscopically behave like a boson, or a collection
of bosons. If the Hamiltonian is of the bilinear form $H=-B^{\dagger }B=- 
\frac{1}{N}\left( \hat{n}+\sum_{i\neq j}^{N}a_{i}^{\dagger }a_{j}\right) $,
the ground state with $n\ll N$ parafermions is $\left( B^{\dagger }\right)
^{n}\left| {\tt 0}\right\rangle $, i.e., $\hat{n}\left[ \left( B^{\dagger
}\right) ^{n}\left| {\tt 0}\right\rangle \right] \approx n\left[ \left(
B^{\dagger }\right) ^{n}\left| {\tt 0}\right\rangle \right] .$

A separate conclusion that follows from this result is that a low-energy
non-interacting qubit system can {\em naturally simulate the dynamics of
bosons}.

\section{Parafermions from Fermions and Bosons}

\label{making}

As stated in the Introduction, qubits do not exist as fundamental particles.
This means that they are either approximate descriptions (e.g., a spin in
the absence of its spatial degrees of freedom), or have to be prepared by
appropriately combining bosons or fermions. I.e., a qubit can be {\em \
encoded} in terms of bosons or fermions under certain conditions (see also 
\cite{Viola:01}). We consider bosonic or fermionic systems with $2N$
single-particle states. Let $k=1,2,\ldots ,N$ denote all relevant quantum
numbers (including spin, if necessary). The following three cases yield
parafermions.

{\em Case 1: Fermionic particle-particle pairs} --- Under the condition $
n_{2k-1}^{f}=n_{2k}^{f}$ it can be shown that $\{f_{2k}f_{2k-1},f_{2k-1}^{
\dagger }f_{2k}^{\dagger }\}=1$ and $[f_{2k-1}f_{2k},f_{2l-1}^{\dagger
}f_{2l}^{\dagger }]=0$ for $k\neq l.$ Furthermore, the set $
\{f_{2k-1},f_{2k-1}^{\dagger }f_{2k}^{\dagger },n_{2k-1}^{f}+n_{2k}^{f}-1\}$
satisfies the commutation relations of $sl(2)$. Therefore the mapping $
a_{k}\Longleftrightarrow f_{2k}f_{2k-1},$ $a_{k}^{\dagger
}\Longleftrightarrow f_{2k-1}^{\dagger }f_{2k}^{\dagger }$ and $
2n_{k}\Longleftrightarrow n_{2k-1}^{f}+n_{2k}^{f}$, is a mapping to
parafermions. The vacuum state of parafermions in this case corresponds to
the vacuum state $|{\tt 0}\rangle _{f}$ of fermions. Example: Cooper pairs.

{\em Case 2: Fermionic particle-hole pairs} --- Under the condition $
n_{2k-1}^{f}+n_{2k}^{f}=1$ it can be shown as in Case 1 that $
a_{k}\Longleftrightarrow f_{2k}^{\dagger }f_{2k-1},$ $a_{k}^{\dagger
}\Longleftrightarrow f_{2k-1}^{\dagger }f_{2k}$ and $2n_{k}-1
\Longleftrightarrow n_{2k-1}^{f}-n_{2k}^{f}$ is a mapping to parafermions.
However, in this case the vacuum state of parafermions is $\left| {\tt 0}
\right\rangle =f_{2N}^{\dagger }\cdots f_{4}^{\dagger }f_{2}^{\dagger
}\left| {\tt 0}\right\rangle _{f}$, because then $a_{k}\left| {\tt 0}
\right\rangle =0$ for all $k$. This vacuum state plays the role of a Fermi
level. Example: excitons. In fact, all quantum computer proposals that use
electrons, e.g., quantum dots \cite{Loss:98}, electrons on Helium \cite{Platzman:99,Dykman:00}, are equivalent to this case. E.g., $f_{2}^{\dagger
}f_{1}$ and $f_{1}^{\dagger }f_{2}$ can represent the transition operators
between two spin states in the quantum dot proposal.

{\em Case 3: Bosonic `particle-hole' pairs} --- Under the condition $
n_{2k-1}^{b}+n_{2k}^{b}=1$ it can be shown as in Case 1 that $
a_{k}\Longleftrightarrow b_{2k}^{\dagger }b_{2k-1},$ $a_{k}^{\dagger
}\Longleftrightarrow b_{2k-1}^{\dagger }b_{2k}$ and $2n_{k}-1
\Longleftrightarrow n_{2k-1}^{b}-n_{2k}^{b}$ is a mapping to parafermions.
However, in this case the vacuum state of parafermions is $\left| {\tt 0}
\right\rangle =b_{2N}^{\dagger }\cdots b_{2k}^{\dagger }\cdots
b_{4}^{\dagger }b_{2}^{\dagger }\left| {\tt 0}\right\rangle _{b}$, again
because then $a_{k}\left| {\tt 0}\right\rangle =0$ for all $k$. Example:
dual-rail photons in the optical quantum computer proposal \cite{Knill:00}.

This classification illustrates the by-necessity compound nature of a qubit,
and puts into a unified context the many different proposals for
constructing qubits in physical systems. Note that it is possible to use
more than two fermions or bosons to construct a parafermion. Further
implications, especially as related to the simulation of models of
superconductivity (Case 1) on a quantum computer, have been explored in Ref.~
\onlinecite{WuByrdLidar:01}.

\section{Parafermionic Bilinear Hamiltonians are Universal but Fermionic and
Bosonic are Not}

\label{bilinuniv}

In this section we discuss an apparently striking difference between the
universality of bilinear Hamiltonians acting on fermions and bosons, as
compared to parafermions. Let us consider the set of
particle-number-conserving bilinear operators of bosons, fermions and
parafermions:

\[
b_{i}^{\dagger }b_{j},\quad f_{i}^{\dagger }f_{j},\quad a_{i}^{\dagger
}a_{j}. 
\]
As noted in Table~\ref{tab1}, in the fermionic case these operators generate
the group $U(N)$ where $N$ is the number of particles. The same is true for
bosons \cite{Wybourne:book}. Clearly, therefore, fermionic and bosonic
Hamiltonians containing only these operators are not universal with respect
to an interesting (i.e., exponentially large) $SU(2^N)$ subgroup. On the
other hand, as discussed in the previous section, these fermionic and
bosonic operators can be used to define parafermionic operators $
a_{i}^{\dagger }a_{j}$ in two-to-one correspondence. As mentioned in Section 
\ref{subalgebras}, the set $\{a_{i}^{\dagger }a_{j}\}_{i,j=1}^{N+1}$
generates the subalgebra SA$n$($N$), with dimension $\frac{(2N)!}{N!N!}
(>2^{N})$ (recall that the total number of $Q_{\alpha ,\beta }(N)$ operators
is $2^{2N}$). The corresponding Lie group appears to be large enough to be
interesting for universal quantum computation. This expectation is borne
out, since one can construct an XY model, Eq.~(\ref{eq:xy}) below, using
the set $\{a_{i}^{\dagger }a_{j}\}$. As shown in Ref.~
\onlinecite{Bacon:Sydney}, the XY model is by itself universal provided
one uses three physical qubits per {\em encoded qutrit}, together with
nearest-neighbor and next-nearest-neighbor interactions (see also Section 
\ref{encoded-example}). We discuss the XY model in detail in Section \ref
{XY} below. First, however, let us argue qualitatively where the difference
between parafermions (qubits) and fermions, bosons originates from. An
example will illuminate this. For the case of bosons and fermions, $
[b_{1}^{\dagger }b_{2},b_{2}^{\dagger }b_{3}]=b_{1}^{\dagger }b_{3}$ and $
[f_{1}^{\dagger }f_{2}$, $f_{2}^{\dagger }f_{3}]=$ $f_{1}^{\dagger }f_{3}$.
But for parafermions, $[a_{1}^{\dagger }a_{2},a_{2}^{\dagger
}a_{3}]=a_{1}^{\dagger }a_{3}(1-2\hat{n}_{2})$. (An easy way to check this, without explicitly calculating the commutator, is
to use the mapping to fermions: $f_{i}^{\dagger }f_{i+1}\leftrightarrow
a_{i}^{\dagger }a_{i+1}$ and the Jordan-Wigner transformation $
f_{i}\rightarrow a_{i}S_{i}$.) Thus the difference is that {\em bosons and
fermions preserve locality, but parafermions do not.}

Similarly, we can consider additional bilinear operators. For fermions, if
we also have $f_{i}f_{j}$ and $f_{j}^{\dagger }f_{i}^{\dagger }$, the group
is $SO(2N)$, which is too small to be interesting for QC. In fact this is a
model of non-interacting fermions: there exists a canonical transformation
to a sum of quadratic terms each of which acts only on a single mode (see
also Refs.~\onlinecite{Bravyi:00,Terhal:01,Knill:01,Viola:01,Zanardi:01}). For bosons,
if we include $b_{i}b_{j}$ and $b_{j}^{\dagger }b_{i}^{\dagger }$, the group
generated is the $N(2N+1)$-parameter symplectic group $Sp(2N,R)$ which is
non-compact, implying that it has no finite dimensional irreps \cite
{Wybourne:book}. If we further include the set of annihilation\ and creation
operators $b_{i}$,$b_{i}^{\dagger }$ together with the identity operator $I$, the set $\{I,b_{i},b_{i}^{\dagger },b_{i}b_{j},b_{j}^{\dagger
}b_{i}^{\dagger },b_{j}^{\dagger }b_{i}\}$ generates the semidirect-product
group $N(N)\otimes Sp(2N,R)$, where $N(N)$ is the Heisenberg group, with $
(N+1)(2N+1)$ generators (Ref.~\onlinecite[Ch.20]{Wybourne:book}). This is therefore still
too small to be interesting for universal QC. In fact, {\em this is exactly
the reason why linear optics by itself is insufficient for universal QC}.
The situation does not change even after introduction of the displacement
operators $D_{i}(\alpha )=\exp (\alpha b_{i}^{\dagger }-$ $\alpha ^{\ast
}b_{i})$ \cite{Ralph:01}, since $D_{i}(\alpha )\in N(N)\otimes Sp(2N,R).$

The way to universality [with respect to $SU(2^{N})$] is to introduce
nonlinear operations such as a Kerr nonlinearity \cite{Chuang:95a},
self-interaction \cite{Zanardi:02}, or conditional measurements \cite{Knill:00,Ralph:01}. A Kerr nonlinearity is a two-qubit interaction of the
form $n_{i}^{b}n_{j}^{b}$ (where $i$ and $j$ are different modes), which
directly provides a {\sc CPHASE} gate. To see this, consider a dual-rail
encoding \cite{Chuang:95a}:\ Suppose that one qubit is encoded into $
\left\vert 0\right\rangle =b_{1}^{\dagger }\left\vert {\tt 0}\right\rangle $, $\left\vert 1\right\rangle =b_{2}^{\dagger }\left\vert {\tt 0}
\right\rangle $, while a second qubit is encoded into $\left\vert
0\right\rangle =b_{3}^{\dagger }\left\vert {\tt 0}\right\rangle $, $
\left\vert 1\right\rangle =b_{4}^{\dagger }\left\vert {\tt 0}\right\rangle $
($\left\vert {\tt 0}\right\rangle $ is the vacuum state). The two-qubit
states are 
\begin{eqnarray*}
\left\vert 00\right\rangle  &=&b_{3}^{\dagger }b_{1}^{\dagger }\left\vert 
{\tt 0}\right\rangle ,\quad \left\vert 01\right\rangle =b_{3}^{\dagger
}b_{2}^{\dagger }\left\vert {\tt 0}\right\rangle  \\
\left\vert 10\right\rangle  &=&b_{4}^{\dagger }b_{1}^{\dagger }\left\vert 
{\tt 0}\right\rangle ,\quad \left\vert 11\right\rangle =b_{4}^{\dagger
}b_{2}^{\dagger }\left\vert {\tt 0}\right\rangle .
\end{eqnarray*}
(This is related to {\em Case 3} of section \ref{making}, where we showed
how to make qubits from bosons.) It is then simple to verify that $\exp
(-i\pi n_{2}^{b}n_{4}^{b})$ acts exactly as a {\sc CPHASE}\ gate, i.e., it
is represented by the matrix ${\rm diag}(1,1,1,-1)$ in this two-qubit basis.
Here we wish to point out that a recently introduced alternative to a Kerr
nonlinearity \cite{Zanardi:02}, namely the self-interaction $(n_{i}^{b})^{2}$, is in fact closely related to the Kerr nonlinearity. Thus methods
developed to use one of these non-linear interactions can be transported to
the other. Let us demonstrate this point by giving a simple circuit to show
how one interaction simulates the other. We start with the operator identity

\[
\exp (\phi (a^{\dagger }b-b^{\dagger }a))b^{\dagger }\exp (-\phi (a^{\dagger
}b-b^{\dagger }a))=\cos \phi b^{\dagger }+\sin \phi a^{\dagger },
\]
which can be proved directly from the Baker-Hausdorff formula
\begin{equation}
e^{-\alpha A}Be^{\alpha A}=B-\alpha \lbrack A,B]+\frac{\alpha ^{2}}{2!}
[A,[A,B]]+\frac{\alpha ^{3}}{3!}[A,[A,[A,B]]]+...  \label{eq:BH}
\end{equation}
Using the latter identity it is then simple to verify the following
identity, {\em which holds on the two-qubit subspace above},

\[
\exp (-i\pi n_{2}^{b}n_{4}^{b})=\exp (-\frac{\pi }{4}(b_{2}^{\dagger
}b_{4}-b_{4}^{\dagger }b_{2}))\exp (-i\pi \frac{
(n_{2}^{b})^{2}+(n_{4}^{b})^{2}-n_{2}^{b}-n_{4}^{b}}{2})\exp (\frac{\pi }{4}
(b_{2}^{\dagger }b_{4}-b_{4}^{\dagger }b_{2})).
\]
This is an exact 3-gate simulation of the Kerr {\sc CPHASE} gate in terms of
the self-interaction. The simulation uses the linear bosonic operators $
b_{i}^{\dagger }b_{j}$ and the local energies $n_{i}^{b}$ in order to
unitarily rotate the self-interaction terms $(n_{2}^{b})^{2}+(n_{4}^{b})^{2}$
to a Kerr interaction.

\section{Fluctuations in Parafermion Number at Finite Temperature}

\label{fluctuations}

So far we have not really made use of the full power of the Fock space
representation, which allows to consider the case of fluctuating particle
number. The quantum statistics of parafermions is determined by their
commutation relations, like fermions (Fermi-Dirac statistics) and bosons
(Bose-Einstein statistics). A simple case to consider is that of
non-interacting parafermions. The Fermi-Dirac distribution for an ideal
Fermi gas is derived using only the restriction that no more than a single
fermion can occupy a given mode \cite{LL:SP1}. Hence the statistics of
non-interacting parafermions is clearly the same as that of non-interacting
fermions.

Fluctuations in particle number will be a result of interaction of the
system with an external bath, which imposes a chemical potential $\mu $
(essentially the gradient of the particle flow). As a simple example,
consider the following system-bath interaction Hamiltonian: 
\begin{equation}
H_{I}=\sum_{i=1}^{N}\sigma _{i}^{z}\otimes B_{i}^{z}\rightarrow
\sum_{i=1}^{N}(2\hat{n}_{i}-1)\otimes B_{i}^{z},  \label{eq:inter}
\end{equation}
where $B_{i}^{z}$ are bath operators. To further simplify things assume the
bath is treated classically, i.e., $B_{i}^{z}$ are positive $c$-numbers.
With this Hamiltonian, one can study the fluctuations of parafermions under
finite temperature $T$. Mapping from the well-know result for a
non-interacting Fermi gas \cite{LL:SP1} it then follows that the average
occupation for the $i^{{\rm th}}$ qubit site is 
\[
\langle n_{i}\rangle =\frac{1}{e^{(2B_{i}^{z}-\mu )/kT}+1} 
\]
where $k$ is Boltzman's constant. This is the average value of the
qubit-``spin'' (whether it is $|0\rangle $ or $|1\rangle $). Keeping the
chemical potential $\mu $ fixed, in the limit of $T\rightarrow 0$ we find
that $\langle n_{i}\rangle \rightarrow 1$ if $B_{i}^{z}<\mu $, but $\langle
n_{i}\rangle \rightarrow 0$ if $B_{i}^{z}>\mu $. Thus, as expected, it is
essential to keep the interaction with the bath weak (compared to $\mu $) to
prevent fluctuations in qubit ``orientation'' at low temperatures. At finite 
$T$ we find $\langle n_{i}\rangle <1$, meaning that some fluctuation is
unpreventable. Of course, our model is very naive, and the picture is
modified when qubit interactions are taken into account. However, it should
be clear that a Fock space description of qubits, i.e., in terms of
parafermions, could be valuable in studying qubit statistics at finite
temperatures.

\section{Universality of Exchange-Type Hamiltonians}

\label{universality}

In this final section we conclude with an application of the formalism we
developed above to the study of the universality power of Hamiltonians. We
have considered this question in detail before for general exchange-type
Hamiltonians (isotropic and anisotropic)\cite{WuLidar:01,LidarWu:01}. We
first briefly review the universality classification of various physically
relevant bilinear Hamiltonians. It will be seen that while in certain cases
the Hamiltonian is not sufficiently powerful to be universal with respect to 
$U(2^{N})$, it is universal with respect to a subgroup. As mentioned in
Section \ref{subalgebras}, this result requires the use of {\em encoding of
physical qubits into logical qubits} \cite
{Bacon:99a,DiVincenzo:00a,Bacon:Sydney,Lidar:00b}. We then consider in
detail the representative example of the XY model, where we give a new
result about universality (in fact, the lack thereof) in the case of
nearest-neighbor-only interactions. We then present new results about codes
with higher rates than considered in Refs.~\onlinecite{WuLidar:01,LidarWu:01}. For simplicity we revert when convenient to the Pauli matrix notation in
this section, which is more familiar to practitioners of QC.

\subsection{Classification of Bilinear Hamiltonians}

The most general bilinear Hamiltonian for a qubit system is

\begin{equation}
H(t)\equiv H_{0}+V+F=\sum_{i}\frac{1}{2}\varepsilon _{i}\sigma
_{i}^{z}+\sum_{i<j}V_{ij}+F,  \label{eq:H(t)}
\end{equation}
where $H_{0}$ is the qubit energy term, the interaction between qubits $i$
and $j$ is: 
\[
V_{ij}=\sum_{\alpha ,\beta =x,y,z}J_{ij}^{\alpha \beta }(t)\sigma
_{i}^{\alpha }\sigma _{j}^{\beta },
\]
and the external single-qubit operations are: 
\[
F=\sum_{i}f_{i}^{x}(t)\sigma _{i}^{x}+f_{i}^{y}(t)\sigma _{i}^{y}.
\]
Recall the \textquotedblleft standard\textquotedblright\ result about
universal quantum computation: The group $U(2^{N})$ on $N$ qubits can be
generated using arbitrary single qubit gates and a non-trivial two-qubit
entangling gate such as {\sc CNOT} \cite{Barenco:95a}. The general
Hamiltonian $H(t)$ can generate such a universal gate set, e.g., as follows:
Suppose there are controllable $\sigma _{i}^{z}$ and $\sigma _{i}^{x}$
terms. Then $\sigma _{i}^{y}$ can be generated using Euler angles: 
\[
\sigma _{i}^{y}=\exp (-i\pi \sigma _{i}^{z}/4)\sigma _{i}^{x}\exp (i\pi
\sigma _{i}^{z}/4).
\]
This is an instance of a simple but extremely useful result: let $A$ and $B$
be anticommuting hermitian operators where $A^{2}=I$ ($I$ is the identity
matrix). Then, using $Ue^{V}U^{\dagger }=e^{UVU^{\dagger }}$ ($U$ is
unitary, $V$ is arbitrary): 
\begin{eqnarray}
C_{A}^{\varphi }\circ \exp (i\theta B) &\equiv &\exp (-iA\varphi )\exp
(i\theta B)\exp (iA\varphi )  \nonumber \\
&=&\left\{ 
\begin{array}{r}
\exp (-i\theta B)\quad {\rm if\,\,}\varphi =\pi /2 \\ 
\exp [i\theta (iAB)]\quad {\rm if\,\,}\varphi =\pi /4
\end{array}
\right. .  \label{eq:C}
\end{eqnarray}
One can also derive these relations for $su(2)\ $angular momentum operators,
without assuming that $\{A,B\}=0$ and $A^{2}=I$. Let $J_{x}$ and $J_{z}$ be
generators of $su(2)$. Then, using the Baker-Hausdorff relation Eq.~(\ref
{eq:BH}), and $[J_{z},J_{x}]=iJ_{y}$:
\[
\exp (-i\varphi J_{z})J_{x}\exp (i\varphi J_{z})=J_{x}\cos \varphi
+J_{y}\sin \varphi .
\]
From here follows, using $Ue^{V}U^{\dagger }=e^{UVU^{\dagger }}$ again:
\[
C_{J_{z}}^{\varphi }\circ \exp (i\theta J_{x})=\exp (i\theta (J_{x}\cos
\varphi +J_{y}\sin \varphi )),
\]
and Eq.~(\ref{eq:C}) can be verified, with $\varphi \rightarrow 2\varphi $.

Different QC proposals usually have different two-qubit interactions.
Typical types include $\sigma _{i}^{z}\sigma _{i+1}^{z},\sigma
_{i}^{y}\sigma _{i+1}^{y}$ $($or $\sigma _{i}^{x}\sigma _{i+1}^{x}),$ $
\sigma _{i}^{x}\sigma _{i+1}^{x}+\sigma _{i}^{y}\sigma _{i+1}^{y}$ (XY
model), and $\overrightarrow{\sigma }_{i}\cdot \overrightarrow{\sigma }_{j}$
(Heisenberg model)$.$ It is simple to show that they can all be transformed
into a common canonical form $\sigma _{i}^{z}\sigma _{i+1}^{z}$, using a few
unitary transformation. The term $\sigma _{i}^{z}\sigma _{i+1}^{z}$ can be
used to generate {\sc CPHASE} and from there, {\sc CNOT} \cite
{Loss:98,Nielsen:book}. E.g., the XY term can first be transformed into $
\sigma _{i}^{x}\sigma _{i+1}^{x}$ using Euler angle rotations about $\sigma
_{i}^{x} $, which flips the sign of the $\sigma _{i}^{y}\sigma _{i+1}^{y}$
term:

\begin{eqnarray*}
\exp \left[ \frac{i\theta }{2}(\sigma _{i}^{x}\sigma _{i+1}^{x}+\sigma
_{i}^{y}\sigma _{i+1}^{y})\right] \left( C_{\sigma _{i}^{x}}^{\pi /2}\circ
\exp \left[ \frac{i\theta }{2}(\sigma _{i}^{x}\sigma _{i+1}^{x}+\sigma
_{i}^{y}\sigma _{i+1}^{y})\right] \right)  \nonumber \\
=\exp (i\theta \sigma _{i}^{x}\sigma _{i+1}^{x}),
\end{eqnarray*}
which can subsequently be transformed into the canonical form using another
Euler angle rotation:

\[
C_{\sigma _{i}^{y}+\sigma _{i+1}^{y}}^{\pi /4}\circ \sigma _{i}^{x}\sigma
_{i+1}^{x}=\sigma _{i}^{z}\sigma _{i+1}^{z}, 
\]
where using $[\sigma _{i}^{y},\sigma _{i+1}^{y}]=0$ we have abbreviated $
C_{\sigma _{i+1}^{y}}^{\pi /4}\circ C_{\sigma _{i}^{y}}^{\pi /4}$ as $
C_{\sigma _{i}^{y}+\sigma _{i+1}^{y}}^{\pi /4}$. The method of Euler angle
rotations as applied here is also known as ``selective recoupling'' in the
NMR literature \cite{Slichter:book}.

Not all QC proposals have an interaction Hamiltonian that appears to be of
the form $V_{ij}$. E.g., the ion-trap proposal \cite{Cirac:95} looks quite
different since it involves interactions between ions mediated by a phonon.
The interaction between the $i^{{\rm th}}$ ion and the phonon has the form $
\sigma _{i}^{-}b^{\dagger }+\sigma _{i}^{+}b$. This is nevertheless
equivalent to an XY model, since: 
\[
\sigma _{i}^{x}\sigma _{i+1}^{x}+\sigma _{i}^{y}\sigma _{i+1}^{y} =
C_{\sigma_i^z-\sigma_{i+1}^z}^{\pi/4} \circ 2i[\sigma _{i}^{-}b^{\dagger
}+\sigma _{i}^{+}b,\sigma _{i+1}^{-}b^{\dagger }+\sigma _{i+1}^{+}b]. 
\]
Therefore in many cases it suffices to study the interaction $\sigma
_{i}^{z}\sigma _{i+1}^{z}.$

Let us now consider a number of more restricted models:

\subsubsection{No external single-qubit operations}

If $F=0$ then the {\em nearest-neighbor} set $\left\{ \sigma _{i}^{z},\sigma
_{i}^{z}\sigma _{i+1}^{z},\sigma _{i}^{x}\sigma _{i+1}^{z},\sigma
_{i+1}^{x}\sigma _{i}^{z}\right\} $ is still universal, since 
\[
\sigma _{i}^{y}=C_{\sigma _{i+1}^{z}\sigma _{i}^{z}}^{\pi /4}\circ \sigma
_{i+1}^{z}\sigma _{i}^{x}. 
\]
This is the case when $H_{0}$ is controllable. More physically, the set $
\left\{ \sigma _{i}^{z},\overrightarrow{\sigma }_{i}\cdot \overrightarrow{
\sigma }_{i+1},(\overrightarrow{\sigma }_{i}\times \overrightarrow{\sigma }
_{i+1})_{y}=\sigma _{i}^{z}\sigma _{i+1}^{x}-\sigma _{i+1}^{z}\sigma
_{i}^{x}\right\} $ is also universal, where $\overrightarrow{ \sigma }
=(\sigma^x,\sigma^y,\sigma^z)$. The term $\overrightarrow{\sigma }
_{i}\times \overrightarrow{\sigma }_{i+1}$ is an anisotropic
(Dzyaloshinskii-Moriya) interaction which arises, e.g., in quantum dots in
the presence of spin-orbit coupling \cite{WuLidar:01,Kavokin:01,Bonesteel:01,Burkard:01,WuLidar:02}.

\subsubsection{No external single-qubit operations and $H_{0}$ uncontrollable
}

If $F=0$ and $H_{0}$ is not controllable, then the nearest-neighbor set $
\left\{ \sigma _{i}^{z}\sigma _{i+1}^{z},\sigma _{i}^{x}\sigma
_{i+1}^{z},\sigma _{i}^{z}\sigma _{i+1}^{x},\sigma _{i}^{y}\sigma
_{i+1}^{z},\sigma _{i}^{z}\sigma _{i+1}^{y}\right\} $ is universal, meaning
that the interaction term $V$ by itself is universal. One way to see this is
to map the set to parafermionic operators and note that it overlaps with the
set that generates the parafermionic $U(2^{N})$ (Table~\ref{tab2}).

\subsubsection{Scalar anisotropic exchange-type interactions}

Consider the case $J_{ij}^{\alpha \beta }=J_{ij}^{\alpha }\delta _{\alpha
\beta }$ (denoting $V$ by $V^{\prime }$), which amounts to limiting the
Hamiltonian to scalar anisotropic exchange-type interactions. Using Eq.~(\ref
{eq:sl2}) we then arrive at the second-quantized form

\begin{eqnarray}
H_{0} &=&\sum_{i}\eta _{i}n_{i}  \nonumber \\
F &=&\sum_{i}\left( f_{i}^{\ast }a_{i}+f_{i}a_{i}^{\dagger }\right) 
\nonumber \\
V^{\prime } &=&\sum_{i<j}\Delta _{ij}(a_{i}a_{j}+a_{i}^{\dagger
}a_{j}^{\dagger })+J_{ij}(a_{i}^{\dagger }a_{j}+a_{j}^{\dagger
}a_{i})+4J_{ij}^{z}n_{i}n_{j}  \label{eq:H0FV'}
\end{eqnarray}
where 
\[
\eta _{i}=\varepsilon _{i}+\left( \sum_{j}J_{ij}^{z}+J_{ji}^{z}\right),
\quad f_{i}=(f_{i}^{x}-if_{i}^{y}), 
\]
\[
\Delta _{ij}=J_{ij}^{x}-J_{ij}^{y},\quad J_{ij}=J_{ij}^{x}+J_{ij}^{y}, 
\]
and we dropped a constant energy term.

$V^{\prime }$ is the so-called XYZ model of solid-state physics.
Considering the structure of $V^{\prime }$ and the classification of
operator algebras we carried out in Sections~\ref{subalgebras},\ref{fermpara}, it should be clear that some immediate conclusions can be drawn about the
universality power of this Hamiltonian. The full Hamiltonian $
H_{0}+V^{\prime }+F$ contains the generators of the parafermionic $ U(2^{N})$
(Table~\ref{tab2}), so it is universal. On the other hand, without external
single qubit operations $F=0$, whence $[H_{0}+V^{\prime },\hat{p}]=0$, so $
H_{0}+V^{\prime }\in $SA$p$, i.e., preserves parity. This immediately
implies that the XYZ model (even with $H_{0})$ is by itself not universal.
However, it can be made universal by {\em encoding} logical qubits into
several (two are in fact sufficient) physical qubits \cite{WuLidar:01}. The
elimination of single qubit operations ($F=0$) can be quite useful, since
typically single and two-qubit operations involve very different
constraints. In some cases single-qubit operations can be very difficult to
implement (see \cite{WuLidar:01,LidarWu:01,DiVincenzo:00a} and references
therein for extensive discussions of this point).

\subsection{XY Model}

\label{XY}

Consider now the XY model, which is defined by 
\begin{equation}
V_{XY}=\sum_{i<j}J_{ij}(a_{i}^{\dagger }a_{j}+a_{j}^{\dagger }a_{i}).
\label{eq:xy}
\end{equation}
It is relevant to a number of proposals for quantum computing, including
quantum Hall systems \cite{Privman:98,Mozyrsky:01}, quantum dots in
microcavities \cite{Imamoglu:99}, quantum dots coupled by exciton exchange 
\cite{Quiroga:99}, and atoms in microcavities \cite{Zheng:00}. Let us
summarize what is currently known about quantum computational universality
of this model.

\begin{itemize}
\item In Ref.~\onlinecite{Imamoglu:99} it was shown that the XY model with
nearest neighbor interactions only, together with single qubit operations,
is universal.

\item In Ref.~\onlinecite{Bacon:Sydney} it was argued that the XY model is
universal without single qubit operations, provided these gates can be
applied between nearest-neighbor and next-nearest-neighbor pairs of qubits.
This involved encoding a logical qutrit into three physical qubits: $
|0_{L}\rangle =|001\rangle $, $|1_{L}\rangle =|010\rangle $, $|2_{L}\rangle
=|100\rangle $. We reconsider this in subsection \ref{highrates} below in
the context of the XXZ model (but using the methods of \cite{LidarWu:01}
the results are valid also for the XY model).

\item In Ref.~\onlinecite{LidarWu:01} we showed that the XY model is
universal using only nearest and next-nearest neighbor ($J_{i,i+2}$)
interactions, together with single qubit $\sigma _{z}$ terms. This too
involved an encoding, of a logical qubit into two physical qubits: $
|0_{L}\rangle =|01\rangle $, $ |1_{L}\rangle =|10\rangle $. Two comments are
in order about this result:\ first, next-nearest neighbor interactions can
be nearest neighbor in 2D (e.g., in an hexagonal array); second, unlike \cite
{Imamoglu:99}, we did not assume the $\sigma _{z}$ terms to be controllable,
i.e., there is no individual control over $\varepsilon _{i}$
[Eq.~(\ref{eq:H(t)})]. A similar model is treated in subsection \ref{xy+yx} below.
\end{itemize}

The question now arises: {\em Is the }XY {\em model universal with
nearest-neighbor interactions only?} We prove that it is not.

The nearest-neighbor XY model in its parafermionic form is 
\[
H=\sum_{i}^{N}\epsilon _{i}n_{i}+\sum_{i}^{N}J_{i,i+1}(a_{i}^{\dagger
}a_{i+1}+a_{i+1}^{\dagger }a_{i}). 
\]
Consulting Table~\ref{tab2}, we see that $H$ can only generate $SU(N)$,
which is clearly too small even for encoded quantum computation.

\subsection{Antisymmetric XY Model}

\label{xy+yx}

To illustrate the idea of encoding for universality, let us briefly consider
the ``antisymmetric XY model'': 
\begin{equation}
V_{{\rm a}XY}=\sum_{i<j}J_{ij}^{xy}\sigma _{i}^{x}\sigma
_{j}^{y}+J_{ij}^{yx}\sigma _{i}^{y}\sigma _{j}^{x}.
\end{equation}
Here $J_{ij}^{xy}$ and $J_{ij}^{yx}$ are real. We encode a logical qubit
into pairs of nearest-neighbor physical qubits. Letting 
\begin{equation}
\widetilde{\Delta }_{ij}=J_{ij}^{xy}-J_{ij}^{yx},\quad \widetilde{J}
_{ij}\equiv J_{ij}^{xy}+J_{ij}^{yx},\quad \epsilon _{m}^{\pm }\equiv
\varepsilon _{2m-1}-\varepsilon _{2m},
\end{equation}
using the compact notation $\cdot _{m}\equiv \cdot _{2m-1,2m}$, and assuming
that interactions are on only inside pairs of qubits encoding one qubit, we
find for the Hamiltonian $H=H_{0}+V_{{\rm a}XY}$: 
\begin{equation}
H_{{\rm a}XY}=\sum_{m=1}^{N/2}\left( \widetilde{J}_{m}R_{m}^{y}+\epsilon
_{m}^{+}R_{m}^{z}\right) +\left( \widetilde{\Delta }_{m}T_{m}^{y}+\epsilon
_{m}^{-}T_{m}^{z}\right) ,
\end{equation}
where the $T$ and $R$ operators were defined in Eqs.~(\ref{eq:slt}),(\ref{eq:slr}). Since the $T$ and $R$ operators form commuting $sl(2)$ algebras,
the Hilbert space splits into two independent computational subspaces. The $
R $ operators conserve parity, so that an appropriate encoding in the
axially symmetric case ($\widetilde{\Delta }_{m}=0$), using standard qubit
notation, is $|0_{L}\rangle =|00\rangle $ and $|1_{L}\rangle =|11\rangle $.
On the other hand, the $T$ operators preserve particle number, so that if $ 
\widetilde{J}_{m}=0$ (axially antisymmetric case) the encoding is $
|0_{L}\rangle =|01\rangle $, $|1_{L}\rangle =|10\rangle $. In both cases
control over the pair of parameters $\{\widetilde{J}_{m},\epsilon _{m}^{+}\}$
(or $\{\widetilde{\Delta }_{m},\epsilon _{m}^{-}\}$) is sufficient for the
implementation of the single-encoded-qubit $SU_{m}(2)$ group (the subscript $
m$ refers to the $m^{{\rm th}}$ logical/encoded qubit).

Logic operations between encoded qubits require the ``encoded selective
recoupling'' method introduced in Ref.~\onlinecite{LidarWu:01}. Consider the
``axially antisymmetric qubit'' $|0_{L}\rangle =|01\rangle $, $|1_{L}\rangle
=|10\rangle $. First, note that using Eq.~(\ref{eq:C}): 
\begin{equation}
C_{T_{12}^{x}}^{\pi /2}\circ T_{23}^{x}=i\sigma _{1}^{z}\sigma
_{2}^{z}T_{13}^{x}.
\end{equation}
Now assume we can control $\widetilde{\Delta }_{13}$; then: 
\begin{equation}
C_{T_{13}^{x}}^{\pi /4}\circ \left( C_{T_{12}^{x}}\circ T_{23}^{x}\right)
=\sigma _{2}^{z}(\sigma _{3}^{z}-\sigma _{1}^{z})/2.
\end{equation}
Since $\sigma _{1}^{z}\sigma _{2}^{z}$ is constant on the code subspace it
can be ignored. On the other hand, $\sigma _{2}^{z}\sigma _{3}^{z}$ acts as $
-T_{1}^{z}T_{2}^{z}$:

\begin{equation}
|0_{L}\rangle _{1}|0_{L}\rangle _{2}=|01\rangle _{12}|01\rangle _{34} 
\stackrel{\sigma _{2}^{z}\sigma _{3}^{z}}{\rightarrow }-|01\rangle
_{12}|01\rangle _{34}=-|0_{L}\rangle _{1}|0_{L}\rangle _{2},
\end{equation}
and similarly for the other three combinations: $|0_{L}\rangle |1_{L}\rangle
\rightarrow |0_{L}\rangle |1_{L}\rangle $, $|1_{L}\rangle |0_{L}\rangle
\rightarrow |1_{L}\rangle |0_{L}\rangle $, $|1_{L}\rangle |1_{L}\rangle
\rightarrow -|1_{L}\rangle |1_{L}\rangle $. I.e., $\sigma _{2}^{z}\sigma
_{3}^{z}$ acts as an encoded $\sigma ^{z}\otimes \sigma ^{z}$. This
establishes universal encoded computation in the antisymmetric XY model.

\subsection{Codes with Higher Rates}

\label{highrates}

The encoding of one logical qubit into two physical qubits is not very
efficient. Can we do better? I.e., can we perform encoded universal QC on
codes with a rate (no. of logical qubit to no. of physical qubits) that is
greater than $1/2$? We will show how in the case of the XXZ model, defined
as $H=H_{0}+H_{XXZ}$, where 
\[
H_{XXZ}=\sum_{i<j}J_{ij}^{x}(\sigma _{i}^{x}\sigma _{j}^{x}+\sigma
_{i}^{y}\sigma _{j}^{y})+J_{ij}^{z}\sigma _{i}^{z}\sigma _{j}^{z}. 
\]
When surface and interface effects are taken into account, the XY-examples
of QC proposals \cite{Imamoglu:99,Privman:98,Mozyrsky:01,Quiroga:99,Zheng:00}, as well as the Heisenberg examples \cite{Loss:98,Kane:98,Vrijen:00}, are
better described by the axially symmetric XXZ model. Additional sources of
non-zero $J_{ij}^{z}$ in the XY-examples can be second-order effects
(e.g., virtual cavity-photon generation without spin-flips\cite{Imamoglu:99}). A natural XXZ-example is that of electrons on helium \cite{Platzman:99,Dykman:00}.

First, note that the code used in the XY model, $|0_{L}\rangle =|01\rangle$, $|1_{L}\rangle =|10\rangle $, is applicable here as well: $T_{ij}^{x} = 
\frac{1}{2}(\sigma^x_i \sigma^x_j + \sigma^y_i \sigma^y_j)$ preserves
particle-number, and serves as an encoded $\sigma ^{x}$; $\sigma _{i}^{z}$
terms from $H_{0}$ serve as encoded $\sigma ^{z}$, and $\sigma
_{i}^{z}\sigma _{i+1}^{z}$ applied to physical qubits belonging to different
encoded qubits acts as encoded $\sigma ^{z}\otimes \sigma ^{z}$.

In the general encoding case we consider a block of $N$ qubits where
codewords are computational basis states (bitstrings of $0$'s and $1$'s): $
\{q_{\alpha }^{\dagger }(N_{\alpha })\left| {\tt 0}\right\rangle \}_{\alpha
} $, where $\alpha =\{\alpha _{i}\}$ and $\alpha _{i}$ can be $0$ or $1$,
while $N_{\alpha }=0...N$. A code-subspace ${\cal C}(N,n)$ will be defined
by having a fixed number $n$ of $1$'s (i.e., of parafermions). Thus there
are 
\[
d_{N,n}\equiv \dim [{\cal C}(N,n)]={\ {
{{N}  \choose {n}}
} } 
\]
codewords in a subspace. Examples are considered below. Note that these
subspaces are decoherence-free under the process of collective dephasing 
\cite{Duan:98}, and have been analyzed extensively in this context in Ref.~
\onlinecite{Kempe:00}. Figure 1 in Ref.~\onlinecite{Kempe:00} provides a
nice graphical illustration of the $ {\cal C}(N,n)$ subspaces. Since the
decoherence-avoidance properties of the codes we consider here have been
extensively discussed before \cite{Duan:98,Kempe:00}, and even implemented
experimentally \cite{Kwiat:00,Kielpinski:01}, we do not address this issue
here. We further note that Ref.~\onlinecite{Kempe:00} provided an
in-principle proof that universal encoded QC is possible on all subspaces $
{\cal C}(N,n)$ independently. However, this proof had several shortcomings:\
(i) it used a short-time approximation, (ii) it did not make explicit
contact with physically realizable Hamiltonians, (iii), it proceeded by
induction, and thus did not explicitly provide an {\em efficient } algorithm
for universal QC. We remedy all these shortcomings here. i.e., we (i) use
only finite-time operations, (ii) use only the XXZ Hamiltonian, (iii)
provide an efficient algorithm that scales polynomially in $N$.

We need a measure that captures how efficient a ${\cal C}(N,n)$ code is. If
there are $d$ codewords, supported over $N$ $p$-dimensional objects ($ p=2$
is the case of bits), and information is measured in units of $q$, then we
define the rate of the code as 
\[
r(d,p,q)=\frac{\log _{q}d}{\log _{q}p^{N}}. 
\]
The traditional definition for qubits is recovered by setting $p=q=2$, i.e.,
the rate of a code is the ratio of the number of logical qubits $\log _{2}d$
to the number of physical qubits $N$, which in our case becomes: 
\begin{equation}
r=\frac{\log _{2}d_{N,n}}{N}\stackrel{{N\gg 1}}{\longrightarrow }S(\epsilon )
\label{eq:r}
\end{equation}
where $\epsilon \equiv k/N$, 
\[
S(\epsilon )=-\epsilon \log _{2}\epsilon -(1-\epsilon )\log _{2}(1-\epsilon
) 
\]
is the Shannon entropy, and we have used the Stirling formula $\log
x!\approx x\log x-x$. Since $S(\frac{1}{2})=1$ the code has a rate that is
asymptotically unity for the ``symmetric subspace'' ${\cal C}(N,N/2)$, where
the number of $1$'s equals the number of $0$'s in each computational basis
state. However, we will not in fact attempt to encode $\log _{2}d_{N,n}$
logical qubits in the subspace ${\cal C}(N,n)$, since the subspace does not
have a natural tensor product structure. Instead we will consider ${\cal C}
(N,n)$ as a subspace encoding a qu$d$it, where $d=d_{N,n}$. Using the
generalized definition of a rate above, and measuring information in units
of $d$ so that each subspace encodes one unit of information, the rate of
such a code is $r=\frac{\log _{d}d}{\log _{d}2^{N}}$. This, however, exactly
coincides with $r$ of Eq.~(\ref{eq:r}). Therefore we see that the advantage
of working with the symmetric subspace ${\cal C}(N,N/2)$ in the limit of
large $N$ is that its rate approaches unity.

Before embarking on the general analysis, let us note that for an encoding
of one logical qubit into $N$ physical qubits, there is a simple
construction in terms of parafermionic operators: $Q_{\alpha ,\beta }(N)$, $
Q_{\alpha ,\beta }^{\dagger }(N)$, and $[Q_{\alpha ,\beta }^{\dagger }(N),$ $
Q_{\alpha ,\beta }(N)]$ (which is a function of parafermion number), form an 
$ su(2)$ algebra in the basis $\left| 0_{L}\right\rangle =q_{\alpha
}^{\dagger }(N_{\alpha })\left| {\tt 0}\right\rangle $ and $\left|
1_{L}\right\rangle =q_{\beta }^{\dagger }(N-N_{\alpha })\left| {\tt 0}
\right\rangle $. E.g., for $N=2$ there are two cases: the sets $
\{a_{1}a_{2},a_{2}^{\dagger }a_{1}^{\dagger },\hat{n}_{1}+\hat{n}_{2}-1\}$
and $\{a_{1}^{\dagger }a_{2},a_{2}^{\dagger }a_{1},\hat{n}_{1}-\hat{n}_{2}\}$, with corresponding bases $\left| 0_{L}\right\rangle =\left| {\tt 0}
\right\rangle $, $\left| 1_{L}\right\rangle =a_{1}^{\dagger }a_{2}^{\dagger
}\left| {\tt 0} \right\rangle $ and $\left| 0_{L}\right\rangle
=a_{1}^{\dagger }\left| {\tt 0 }\right\rangle $, $\left| 1_{L}\right\rangle
=a_{2}^{\dagger }\left| {\tt 0 } \right\rangle$. These two encodings are
universal (in the sense of blocks of $N$ physical qubits) when only $H_{0}$
and $ V^{\prime }$ are controllable [Eq.~(\ref{eq:H0FV'})].

Let us now move on to the general subspace case, starting with an example.

\subsubsection{Encoded Operations: Example}

\label{encoded-example}

Consider ${\cal C}(3,1)={\rm Span}\{|0\rangle \equiv |001\rangle ,|1\rangle
\equiv |010\rangle ,|2\rangle \equiv |100\rangle \}$, i.e., an encoding of a
logical qutrit into $3$ physical qubits, as in Ref.~\onlinecite{Bacon:Sydney}. Let us count qubits as $i=0,...,N-1$. Our first task is to show how to
generate $su(3)$ on this subspace. It is simple to check that $
T_{01}^{x}|001\rangle =0 $, $ T_{01}^{x}|010\rangle =|100\rangle $, $
T_{01}^{x}|100\rangle =|010\rangle $, and in total

\[
T_{01}^{x}=\left( 
\begin{array}{ccc}
0 & 0 & 0 \\ 
0 & 0 & 1 \\ 
0 & 1 & 0
\end{array}
\right) =|1\rangle \langle 2|+|2\rangle \langle 1|\equiv X_{12} 
\]
where the notation $X_{12}$ denotes a $\sigma ^{x}$ operation between states 
$|1\rangle \equiv |010\rangle $ and $|2\rangle \equiv |100\rangle $.
Similarly it is simple to check that $T_{12}^{x}=X_{01}$ and $
T_{02}^{x}=X_{02}$. Further, using $T_{ij}^{z}\equiv \frac{1}{2}\left(
\sigma _{i}^{z}-\sigma _{j}^{z}\right) $ we have: $T_{01}^{z}=Z_{12}$, $
T_{12}^{z}=Z_{01}$, and $T_{02}^{z}=Z_{02}$, where $Z_{12}$ denotes a $
\sigma ^{z}$ operation between states $|1\rangle $ and $|2\rangle $, etc.
Therefore each pair $\{T_{ij}^{x},T_{ij}^{z}\}$ generates an encoded $su(2)$. But in the sense of generating, $su(N)$ is a sum of overlapping $su(2)$'s 
\cite{Cahn:book}, so using just the nearest neighbor interactions $
\{T_{01}^{x},T_{01}^{z},T_{12}^{x},T_{12}^{x}\}$ we can generate all of $
su(3)$ on ${\cal C}(3,1)$. Note that $[X_{01},X_{12}]=iY_{02}$, so that $
su(2)$ between states $|0\rangle ,|2\rangle $ can in fact be generated using 
$T_{ij}^{x}$'s alone, without $T_{ij}^{z}$'s. This conclusion clearly holds
for the generation of all of $su(3)$ on ${\cal C}(3,1)$, as first pointed
out in Ref.~\onlinecite{Bacon:Sydney}.

Next, we need to show how to implement encoded logical operations between
two ${\cal C}(3,1)$ code subspaces. Let us number the qubits as $i=0,1,2$
for the first block, $i=3,4,5$ for the second block. Consider the effect of
turning on $J_{23}^{z}$, i.e., consider the action of $\sigma _{2}^{z}\sigma
_{3}^{z}$ on the tensor product space ${\cal C}(3,1)\otimes {\cal C}(3,1)$.
The operator $\sigma _{2}^{z}\sigma _{3}^{z}$ is represented by a $9$
-dimensional diagonal matrix on this space, which is easily found to have
the following form in the ordered basis $\{|0\rangle \otimes |0\rangle
,|0\rangle \otimes |1\rangle ,...,|2\rangle \otimes |2\rangle \}$: 
\[
\sigma _{2}^{z}\sigma _{3}^{z}={\rm diag}(-1,1,1,-1,1,1,1,-1,-1)={\rm \ diag}
(-1,1,1)\otimes {\rm diag}(1,1,-1). 
\]
E.g., $\sigma _{2}^{z}\sigma _{3}^{z}|2\rangle \otimes |2\rangle =\sigma
_{2}^{z}\sigma _{3}^{z}|100\rangle \otimes |100\rangle
=(+|100\rangle)\otimes (-|100\rangle )=-|2\rangle \otimes |2\rangle $, which
explains the $ -1$ in the $9^{{\rm th}}$ position in the diagonal matrix
above. The important point is that $\sigma _{2}^{z}\sigma _{3}^{z}$ acts as
a tensor product operator on ${\cal C}(3,1)\otimes {\cal C}(3,1)$, which 
{\em puts a relative phase} between the basis states of each ${\cal C}(3,1)$
factor. This means that $\sigma _{2}^{z}\sigma _{3}^{z}$ acts as an ``$su(3)$
-like'' $\sigma ^{z}\otimes \sigma ^{z}$ on ${\cal C}(3,1)\otimes {\cal C}
(3,1)$.
(It is an ``$su(3)$-like'' $\sigma ^{z}\otimes \sigma ^{z}$ since for $su(2)$ 
$\sigma ^{z}={\rm diag}(1,-1)$ and here we have instead ${\rm diag}(-1,1,1)$
and ${\rm diag}(1,1,-1)$.) It is well known \cite{Nielsen:book} that the 
{\sc CPHASE} gate can be generated from the Hamiltonian $\sigma ^{z}\otimes
\sigma ^{z}$. The same holds here, so that we can generate a CPHASE gate
between two ${\cal C}(3,1)$ subspaces by simply turning on a
nearest-neighbor interaction between the last qubit in the first block and
the first qubit in the second block.

With this example in mind we can move on to the general case.

\subsubsection{Encoded Operations: General Subspace Case}

Let us now consider the case of a general subspace ${\cal C}(N,n)$. We can
enumerate the codewords as $\{|0\rangle ,...,|d_{N,n}\rangle \}$ where $
|0\rangle =|0,...,01,...,1\rangle $ etc., to $|d_{N,n}\rangle
=|1,...,10,...,0\rangle $, where there are $N$ qubits in total and $n$ $1$'s
in each codeword. Consider a fixed nearest-neighbor pair of qubits at
positions $i,i+1$, and the action of $T_{i,i+1}^{x},T_{i,i+1}^{z}$. The four
possibilities for qubit values at these positions are $\{00,01,10,11\}$. Now
consider a pair of codewords $|t\rangle $,$|t^{\prime }\rangle $ such that $
|t\rangle $ has $01$ in the $i,i+1$ positions while $|t^{\prime }\rangle $
has $10$ in the $i,i+1$ positions, and they are identical everywhere else.
We can always find such a pair by definition of ${\cal C}(N,n)$. The action
of $T_{i,i+1}^{x},T_{i,i+1}^{z}$ on $|t\rangle $,$|t^{\prime }\rangle $ is
to generate $su(2)$ between them, just as shown in the case of ${\cal C}
(3,1) $ above. On the other hand the action of $T_{i,i+1}^{x},T_{i,i+1}^{z}$
in the case of $00$ or $11$ in the $i,i+1$ positions is to annihilate all
corresponding codewords (which are anyhow outside of the given ${\cal C}
(N,n) $ subspace). This null action means that, when exponentiated, $
T_{i,i+1}^{x},T_{i,i+1}^{z}$ act as identity on these codewords. Therefore
the action of $T_{i,i+1}^{x},T_{i,i+1}^{z}$ is precisely to generate $su(2)$
between $|t\rangle $,$|t^{\prime }\rangle $, and nothing more. Denote this
by $su(2)_{i,i+1}^{(1)}$. Let us now keep the $01$ and $10$ at positions $
i,i+1$ fixed, and vary all other $N-2$ positions in $|t\rangle $,$|t^{\prime
}\rangle $, subject to the constraint of $n$ $1$ 's, and in the same manner
in both $|t\rangle $,$|t^{\prime }\rangle $. We then run over $K={\ {
{{N-2}  \choose {n-1}}
} }$ codewords, and $T_{i,i+1}^{x},T_{i,i+1}^{z}$ generate $su(2)$ between
each pair of new $|t\rangle $,$|t^{\prime }\rangle $. Denote these by $
su(2)_{i,i+1}^{(k)}$, $k=1..K$. By further letting $i=0,..N-2$ we generate $
N-1$ {\em \ overlapping} $su(2)$'s. These $su(2)$'s can be connected by
swaps so that we can generate all $su(2)_{i,j}^{(k)}$, $k=1..K$, $i<j$. We
thus have a total of ${\ {
{{N-2}  \choose {n-1}}
} }{\ {
{{N}  \choose {2}}
} }$ $su(2)$'s. To generate the entire $su(d_{N,n})$ we need no more than $
d_{N,n}={\ {
{{N}  \choose {n}}
} }$ overlapping $su(2)$'s. Since ${\ {
{{N-2}  \choose {n-1}}
} }{\ {
{{N}  \choose {2}}
} }/{\ {
{{N}  \choose {n}}
} }=\frac{1}{2}n(N-n)>1$, we have more than enough overlapping $su(2)$'s,
and $ su(d_{N,n})$ can be generated.

What is left is to show that we can perform a controlled operation between
two ${\cal C}(N,n)$ subspaces. To do so we again use the nearest-neighbor
interaction $\sigma _{N-1}^{z}\sigma _{N}^{z}$, where the first factor ($
\sigma _{N-1}^{z}$) acts on the last qubit ($N-1$) of the first ${\cal C}
(N,n)$ subspace, and the second factor ($\sigma _{N}^{z}$) acts on the first
qubit ($N$) of the second ${\cal C}(N,n)$ subspace. Now let us sort the
codewords in the two subspaces in an identical manner, e.g., by increasing
binary value. Then consider the action of $\sigma _{N-1}^{z}\sigma _{N}^{z}$
on the resulting ordered basis $\{|0\rangle \otimes |0\rangle ,|0\rangle
\otimes |1\rangle ,...,|d_{N,n}\rangle \otimes |d_{N,n}\rangle \}$. This
action generates a representation of $\sigma _{N-1}^{z}\sigma _{N}^{z}$ by a 
$d_{N,n}\times d_{N,n}$ diagonal matrix. As in the ${\cal C}(3,1)$ case
considered above, this matrix is actually a tensor product of an ``$
su(d_{N,n})$-like'' $\sigma ^{z}\otimes \sigma ^{z}$ on ${\cal C}
(N,n)\otimes {\cal C}(N,n)$. It is simple to determine the form of these two
(different) $\sigma ^{z}$'s. For the codewords belonging to the left ${\cal 
C }(N,n)$ factor write down a $+1$ ($-1$) for each $0$ ($1$) in the $N^{{\rm 
th }}$ position. These numbers are the diagonal entries of the left ``$
su(d_{N,n})$-like'' $\sigma ^{z}$ factor. Similarly, for the codewords
belonging to the right ${\cal C}(N,n)$ factor write down a $+1$ ($-1$) for
each $0$ ($1$) in the $N+1^{{\rm th}}$ position. These numbers are the
diagonal entries of the right ``$su(d_{N,n})$-like'' $\sigma ^{z}$ factor.
Since each such ``$su(d_{N,n})$-like'' $\sigma ^{z}$ puts relative phases
between the basis states of ${\cal C}(N,n)$, the action of $\sigma
_{N-1}^{z}\sigma _{N}^{z}$ is that of a generalized CPHASE between the two
code subspaces. This is sufficient together with $su(d_{N,n})$ on each block
to perform universal quantum computation \cite{comment7}.

\section{Conclusions}

\label{conclusions}

The standard quantum information-theoretic approach to qubits and operations
on qubits, emphasizes qubits as {\em vectors} in a Hilbert space and
operations as {\em transformations} of these vectors \cite{Nielsen:book}.
This is the point of view of the first-quantized\ formulation of quantum
mechanics. An alternative, mathematically equivalent, point of view is the
Fock space, second-quantized formulation of quantum mechanics, which
emphasizes the particle-like nature of quantum states. Qubit up/down states
are replaced by qubit presence/absence, while rotations are replaced by
operators that count or change particle occupation numbers. The mapping of
qubits to parafermions discussed in this paper is a mapping between these
first and second quantized formulations. It proved to be a useful tool in
studying the connection between qubits, bosons and fermions, in analyzing
the algebraic structure of qubit Hamiltonians, and in studying related
quantum computational universality questions. In particular, it allowed us
to classify subalgebras of fermion, boson, and qubit operators and decide
their power for quantum computational universality. These results are
relevant for physical implementation of quantum computers:\ a physical $N$
-qubit system comes equipped with a given Hamiltonian, which generates a
subalgebra of $su(2^{N})$. It is important to know whether this Hamiltonian
is by itself universal or needs to be supplemented with additional
operations, or whether one needs to encode physical qubits into logical
qubits in order to attain universality. Our classification settles this
question for many subalgebras of physical interest.

Another potential advantage of the parafermionic approach, as a
second-quantized formalism for qubits, lies in its ability to naturally deal
with a ``qubit-field'', i.e., situations where the qubit number is not a
conserved quantity. This is certainly a concern for optical and various
solid-state quantum computer implementations. We leave the study of a qubit
field theory as an open area for future explorations.

\section{Acknowledgements}

This material is based on research sponsored by the Defense
Advanced Research Projects Agency under the QuIST program and managed by the
Air Force Research Laboratory (AFOSR), under agreement F49620-01-1-0468.
The U.S. Government is authorized to reproduce and distribute reprints for
Governmental purposes notwithstanding any copyright notation thereon.  The
views and conclusions contained herein are those of the authors and should
not be interpreted as necessarily representing the official policies or
endorsements, either expressed or implied, of the Air Force Research
Laboratory or the U.S. Government.
D.A.L. further gratefully acknowledges financial support from PREA, NSERC, PRO, and the
Connaught Fund. We thank Dr. M.S. Byrd for useful discussions.

\begin{table}[tbp]
\begin{tabular}{l|l|l}
group & fermions & parafermions \\ \hline
$U(2^{N})$ & $Q_{\alpha ,\beta }^{f}(N)$ & $Q_{\alpha ,\beta }(N)$ \\ 
$SO(2N+1)$ & $f_{i}^{\dagger }f_{j},f_{i}f_{j},f_{i},{\rm h.c.}$ & $
a_{i}^{\dagger }S_{i}S_{j}a_{j},a_{i}S_{i}S_{j}a_{j},a_{i}S_{i},{\rm h.c.} $
\\ 
$SO(2N)$ & $f_{i}^{\dagger }f_{j},f_{i}f_{j},{\rm h.c.}$ & $a_{i}^{\dagger
}S_{i}S_{j}a_{j},a_{i}S_{i}S_{j}a_{j},{\rm h.c.}$ \\ 
$U(N)$ & $f_{i}^{\dagger }f_{j}$ & $a_{i}^{\dagger }S_{i}S_{j}a_{j}$
\end{tabular}
\caption{Infinitesimal generators (h.c.=hermitian conjugate).}
\label{tab1}
\end{table}

\begin{table}[tbp]
\begin{tabular}{l|l|l}
group & fermions & parafermions \\ \hline
$U(2^{N})$ & $f_{i}S_{i}^{f},f_{i}^{\dagger }f_{i+1},{\rm h.c.}$ & $
a_{i},a_{i}^{\dagger }a_{i+1},{\rm h.c.}$ \\ 
$SO(2N+1)$ & $f_{i},{\rm h.c.}$ & $a_{i}S_{i},{\rm h.c.}$ \\ 
$SO(2N)$ & $f_{i}^{\dagger }f_{i+1},f_{i}f_{i+1},{\rm h.c.}$ & $
a_{i}^{\dagger }a_{i+1},a_{i}a_{i+1},{\rm h.c.}$ \\ 
$SU(N)$ & $f_{i}^{\dagger }f_{i+1},{\rm h.c.}$ & $a_{i}^{\dagger }a_{i+1}, 
{\rm h.c.}$
\end{tabular}
\caption{Hamiltonian generators.}
\label{tab2}
\end{table}

\end{document}